\documentclass[preprint,amsmath,amssymb,prd,superscriptaddress]{revtex4}
\usepackage[dvips]{graphicx}
\usepackage{color}
\usepackage{multirow}
\usepackage{array}
\usepackage{amsmath,amssymb,amsfonts,amstext,amsthm}
\usepackage{algorithm,algorithmic}
\newcommand{\Dodwf}{\mathcal{D}}
\newcommand{\p}{\partial}
\newcommand{\pslash}{p\kern-1ex /}
\newcommand{\lslash}{l\kern-1ex /}
\newcommand{\kslash}{k\kern-1ex /}
\newcommand{\dslash}{\p\kern-1.2ex /}
\newcommand{\Dslash}{{\cal D}\kern-1.5ex /}

\newcommand{\tr}{{\rm tr}}

\newcommand{\CH}{{\cal H}}
\newcommand{\bea}{\begin{eqnarray}}
\newcommand{\eea}{\end{eqnarray}}

\newcommand{\nn}{\nonumber\\}

\newcommand{\BAN}{\begin{eqnarray*}}
\newcommand{\EAN}{\end{eqnarray*}}
\begin{document}

\newcommand{\NTU}{
  Department of Physics,
  National Taiwan University, Taipei~10617, Taiwan
}

\newcommand{\CQSE}{
  Center for Quantum Science and Engineering,
  National Taiwan University, Taipei~10617, Taiwan
}

\newcommand{\CTS}{
  Center for Theoretical Sciences,
  National Taiwan University, Taipei~10617, Taiwan
}

\preprint{NTUTH-15-505A}

\title{Domain-Wall Fermion with $ R_5 $ Symmetry}

\author{Ting-Wai~Chiu}
\affiliation{\NTU}
\affiliation{\CQSE}


\vskip 8cm 
\begin{center}
{\it This work is dedicated to late Professor James S. Ball}
\end{center}

\begin{abstract}

We present the domain-wall fermion operator which is reflection symmetric in the fifth dimension, with  
the approximate sign function $ S(H) $ of the effective 4-dimensional Dirac operator
satisfying the bound $ |1-S(\lambda)| \le 2 d_Z $ for $ \lambda^2 \in [\lambda_{min}^2, \lambda_{max}^2] $,  
where $ d_Z $ is the maximum deviation $ | 1- \sqrt{x} R_Z(x) |_{\rm max} $ of the 
Zolotarev optimal rational polynomial $ R_Z(x) $ of $ 1/\sqrt{x} $ for $ x \in [1, \lambda_{max}^2/\lambda_{min}^2] $,  
with degrees $ (n-1,n) $ for $ N_s = 2n $, and $ (n,n) $ for $ N_s = 2 n + 1 $.

\end{abstract}

\maketitle


\section{Introduction}

In lattice QCD with exact chiral symmetry \cite{Kaplan:1992bt, Neuberger:1997fp, Narayanan:1994gw}, 
the overlap Dirac operator with bare quark mass $ m_q $ in general can be written as 
\bea
\label{eq:D_mq} 
D(m_q) = m_q + \frac{(1- r m_q)}{2r} \left[ 1 + \gamma_5 \frac{H}{\sqrt{H^2}} \right], \hspace{2mm}  
r = 1/[2m_0(1-dm_0)], \hspace{2mm} m_0 \in (0,2), 
\eea
where $ H = c H_w (1 + d \gamma_5 H_w)^{-1} $, and $ c $ and $ d $ are constants.
Here $ H_w =\gamma_5 D_w $, and $ D_w $ is the standard Wilson-Dirac operator minus the parameter $ m_0 \in (0,2) $.  

The eigenmodes of (\ref{eq:D_mq}) consist of complex conjugate pairs, and (for topologically non-trivial gauge field) real
eigenmodes with definite chiralities at $ m_q $ and $ 1/r $ satisfying the chirality sum rule \cite{Chiu:1998bh}, 
$ n_+ - n_- + N_+ - N_- = 0 $, 
where $ n_{\pm} $ ($ N_{\pm} $) denote the number of eigenmodes at $ m_q $ ($ 1/r $) with $ \pm $ chirality.
Empirically, the real eigenmodes always satisfy either ($ n_- = N_+ = 0$, $n_+ = N_- $) or ($ n_+ = N_- = 0$, $ n_- = N_+ $).
Thus, we have  
\BAN
\det D(m_q) = \left\{ \begin{array}{ll} 
(r m_q)^{n_+} \det \CH_-^2 = (r m_q)^{-n_+} \det \CH_+^2, & \hspace{2mm} n_+ \geq 0,  \\ 
(r m_q)^{n_-} \det \CH_+^2 = (r m_q)^{-n_-} \det \CH_-^2, & \hspace{2mm} n_- \geq 0,   
                 \end{array}\right. 
\EAN
where $ \CH_\pm^2 = P_\pm (D^\dagger D) $, and $ P_\pm = (1 \pm \gamma_5 )/2 $.  
It follows that the pseudofermion action for any number of flavors of  
overlap fermion can be expressed in terms of $ n_\pm $ and $ \CH_\pm^2 $ (Hermitian and positive-definite), 
thus is amenable to the hybrid Monte Carlo simulation (HMC) \cite{Duane:1987de}.
However, this approach requires the computation of the change of $ n_\pm $ at each step of the molecular dynamics in HMC, 
which is prohibitively expensive for large lattices \cite{Fodor:2003bh,DeGrand:2006ws}.
Moreover, the discontinuity of the fermion determinant at the topological boundary highly suppresses  
the crossing rate between different topological sectors, thus renders HMC failing to sample all topological sectors ergodically.

These difficulties can be circumvented by using domain-wall fermion (DWF) with finite $ N_s $ in the fifth dimension. 
Then HMC of lattice QCD with DWF on the 5-dimensional lattice can sample all topological sectors ergodically 
and also keep the chiral symmetry at a good precision. 
This has been demonstrated for 2-flavors QCD \cite{Chiu:2011dz,Chen:2014hva}, and (1+1)-flavors QCD \cite{Chen:2014hyy}. 
For DWF with finite $ N_s $, it is vital to preserve the chiral symmetry maximally,  
or equivalently, the approximate sign function $ S(H) $ of the effective 4-dimensional Dirac operator
satisfies the bound, $ |1-S(\lambda)| \le d_Z $ for $ \lambda^2 \in [\lambda_{min}^2, \lambda_{max}^2] $, 
where $ d_Z $ is the maximum deviation $ | 1- \sqrt{x} R_Z(x) |_{\rm max} $ of the 
Zolotarev optimal rational approximation $ R_Z(x) $ of $ 1/\sqrt{x} $ for $ x \in [1, \lambda_{max}^2/\lambda_{min}^2] $,  
with degrees $ (n,n) $ for $ N_s = 2n+1 $, and $ (n-1,n) $ for $ N_s = 2 n $.
This can be attained by assigning a weight to each layer (along the fifth dimension) of the DWF, according to the formula 
\cite{Chiu:2002ir}
\bea
\label{eq:omega_s}
\omega_s = \frac{1}{\lambda_{min}} \sqrt{ 1 - \kappa'^2 \mbox{sn}^2
                  \left( v_s ; \kappa' \right) }, \hspace{4mm} s = 1, \cdots, N_s, 
\eea
where $ \mbox{sn}( v_s; \kappa' ) $ is the Jacobian elliptic function
with argument $ v_s $ (see Eq. (13) of Ref. \cite{Chiu:2002ir}) and 
modulus $ \kappa' =\sqrt{ 1 - 1/b } $, $ b = \lambda_{max}^2 / \lambda_{min}^2 $. 
Here $ \lambda_{max}^2 $ and $ \lambda_{min}^2 $ are the upper-bound and lower-bound for the eigenvalues of $ H^2 $. 
It should be emphasized that $ \lambda_{min} $ and $ \lambda_{max} $ have to be fixed properly for  
each set of simulations, depending on the parameters $ \beta = 6/g^2 $, quark masses, and lattice size, 
such that the desired precision of chiral symmetry can be attained with the minimal cost of the simulation. 

Nevertheless, (\ref{eq:omega_s}) breaks the $ R_5 $ (reflection) symmetry in the fifth dimension, which is essential for  
obtaining the exact pseudofermion action for hybrid Monte Carlo simulation of one-flavor DWF \cite{Chen:2014hyy}, as 
well as other applications. In this paper, we obtain the weights satisfying the $ R_5 $ symmetry 
($ \omega_s = \omega_{N_s - s + 1}, \ s = 1, \cdots, N_s $) such that     
the approximate sign function $ S(H) $ of the effective 4-dimensional Dirac operator
satisfies the bound, $ |1-S(\lambda)| \le 2 d_Z $, which is twice of that of  
the optimal DWF without $ R_5 $ symmetry \cite{Chiu:2002ir}. 
In other words, if one imposes the $ R_5 $ symmetry, this is the optimal chiral symmety one can have.

\section{Optimal Domain-Wall Fermion}

In general, the 5-dimensional lattice operator of DWF can be written as \cite{Chen:2012jya}
\bea
\label{eq:D_odwf}
[\Dodwf(m)]_{xx';ss'} &=&
  (\rho_s D_w + 1)_{xx'} \delta_{ss'}
 +(\sigma_s D_w - 1)_{xx'} L_{ss'},
\eea
where $ \rho_s = c \omega_s + d $, $ \sigma_s = c \omega_s - d $, and $ c $, $d$ are constants.
The indices $ x $ and $ x' $ denote the sites on the 4-dimensional space-time lattice,
and $ s $ and $ s' $ the indices in the fifth dimension, while
the lattice spacing $ a $ and the Dirac and color indices have been suppressed.
The operator $ L $ is independent of the gauge field, and it can be written as  
\BAN
L = P_+ L_+ + P_- L_-, \quad P_\pm = (1\pm \gamma_5)/2,
\EAN
and
\BAN
\label{eq:L}
(L_+)_{ss'} = (L_-)_{s's}= \left\{ 
    \begin{array}{ll} 
      - m \delta_{N_s,s'}, & s = 1, \\  
      \delta_{s-1,s'}, & 1 < s \leq N_s,   
    \end{array}\right.
\EAN
where $ N_s $ is the number of sites in the fifth dimension,    
$ m \equiv r m_q $, $m_q $ is the bare quark mass, and $ r = 1/[2m_0(1-dm_0)] $.
Including the action of the Pauli-Villars fields (with bare mass $ m_{PV} = 1/r $),
the partition function of DWF in a gauge background can be integrated successively to obtain 
the fermion determinant of the effective 4-dimensional Dirac operator,   
\BAN
Z = \int [d \Psi] [ d \bar \Psi] [d \Phi] [d \Phi^\dagger]
    \exp \left\{-\bar \Psi \Dodwf(m) \Psi - \Phi^\dagger \Dodwf(1) \Phi \right\} = \det D(m_q), 
\EAN
where
\bea
\label{eq:D_eff}
D(m_q) &=& (D_c + m_q)(1+r D_c)^{-1} = m_q + \frac{(1- r m_q)}{2 r} [1 + \gamma_5 S(H)],  \\
\label{eq:Dc}
D_c &=& \frac{1}{r} \frac{1+ \gamma_5 S(H)}{1-\gamma_5 S(H)}, \nn
\label{eq:SH}
S(H) &=& \frac{1-\prod_{s=1}^{N_s} T_s}{1+\prod_{s=1}^{N_s} T_s}, \hspace{2mm}
T_s = \frac{1-\omega_s H }{1+\omega_s H}, \hspace{2mm} H=c H_w ( 1 + d \gamma_5 H_w)^{-1}, \hspace{2mm} H_w = \gamma_5 D_w.  
\eea

For the optimal DWF without $ R_5 $ symmetry \cite{Chiu:2002ir}, 
the weights $ \{ \omega_s \} $ are fixed according to the formula (\ref{eq:omega_s}),    
then $ S(H) $ is equal to the Zolotarev optimal rational approximation of the sign function of $ H $, i.e, $ S(H) = H R_Z(H^2) $,  
satisfying the bound, $ |1-S(\lambda)| \le d_Z $ for $ \lambda^2 \in [\lambda_{min}^2, \lambda_{max}^2] $, 
where $ d_Z $ is the maximum deviation $ | 1- \sqrt{x} R_Z(x) |_{\rm max} $ of the 
Zolotarev optimal rational approximation $ R_Z(x) $ of $ 1/\sqrt{x} $ for $ x \in [1, \lambda_{max}^2/\lambda_{min}^2] $,  
with degrees $ (n,n) $ for $ N_s = 2n+1 $, and $ (n-1,n) $ for $ N_s = 2 n $.
However, it breaks the $ R_5 $ symmetry, i.e., $ R_5 \Dodwf(m) R_5 \neq \Dodwf(m) $, 
where $ (R_5)_{s,s'} = \delta_{s',N_s+1-s} $.
In the following, we construct the optimal DWF satisfying the $ R_5 $ symmetry.

\section{Optimal Domain-Wall Fermion with $R_5$ Symmetry}

First we recall the basic features of the   
optimal rational approximation of the ratio of two positive and continuous functions $ f(x)/g(x) $ for $ x \in [1, b] $. 
Let $ R^{(m,n)}(x) $ denote an irreducible rational polynomial of the form
\BAN
\label{eq:Rmn}
R^{(m,n)}(x)=
\frac{ p_{m} x^{m} + p_{m-1} x^{m-1} + \cdots + p_0 }
     { q_{n} x^{n} + q_{n-1} x^{n-1} + \cdots + q_0 }, \ ( n \ge m, \ p_i, q_i > 0 ). 
\EAN
According to de la Vall\'{e}e-Poussin's theorem and Chebycheff's theorem,
the necessary and sufficient condition for $ R^{(m,n)}(x) $ to be
the optimal rational approximation of $ f(x)/g(x) $ for $ x \in [1, b] $
is that $ \delta(x) \equiv f(x) - g(x) R^{(m,n)}(x) $ 
has $ (m+n+2) $ alternate change of sign in the interval $ [1, b] $,
and attains its maxima and minima (all with equal magnitude), say,
\BAN
\delta(x) =  +\Delta, -\Delta, \cdots, (-1)^{m+n+1} \Delta
\EAN
at consecutive points ($ x_i, i=1,\cdots, m+n+2 $)
\BAN
1 = x_1 < x_2 < \cdots < x_{m+n+2} = b.
\EAN
For $ f(x)/g(x) = 1/\sqrt{x} $, Zolotarev obtained two  
optimal rational polynomials in 1877, $ R_Z^{(n-1,n)}(x) $ and $ R_Z^{(n,n)}(x) $ \cite{Zol:1877}, 
in terms of the Jacobian elliptic functions.  
The basic formulas of Zolotarev optimal rational polynomial are collected in the Appendix. 
(See also Refs. \cite{Akhiezer:1990, Chiu:2002eh}.)  

For the optimal DWF without $ R_5 $ symmetry \cite{Chiu:2002ir}, 
the approximate sign function $ S(H) $ of the effective 4-dimensional lattice Dirac operator is  
exactly equal to $ H R_Z^{(n-1,n)}(H^2) $ for $ N_s = 2n $, and $ H R_Z^{(n,n)}(H^2) $ for $ N_s = 2n + 1 $.

\begin{figure*}[tb]
\begin{center}
\begin{tabular}{@{}c@{}c@{}}
\includegraphics*[width=8cm,clip=true]{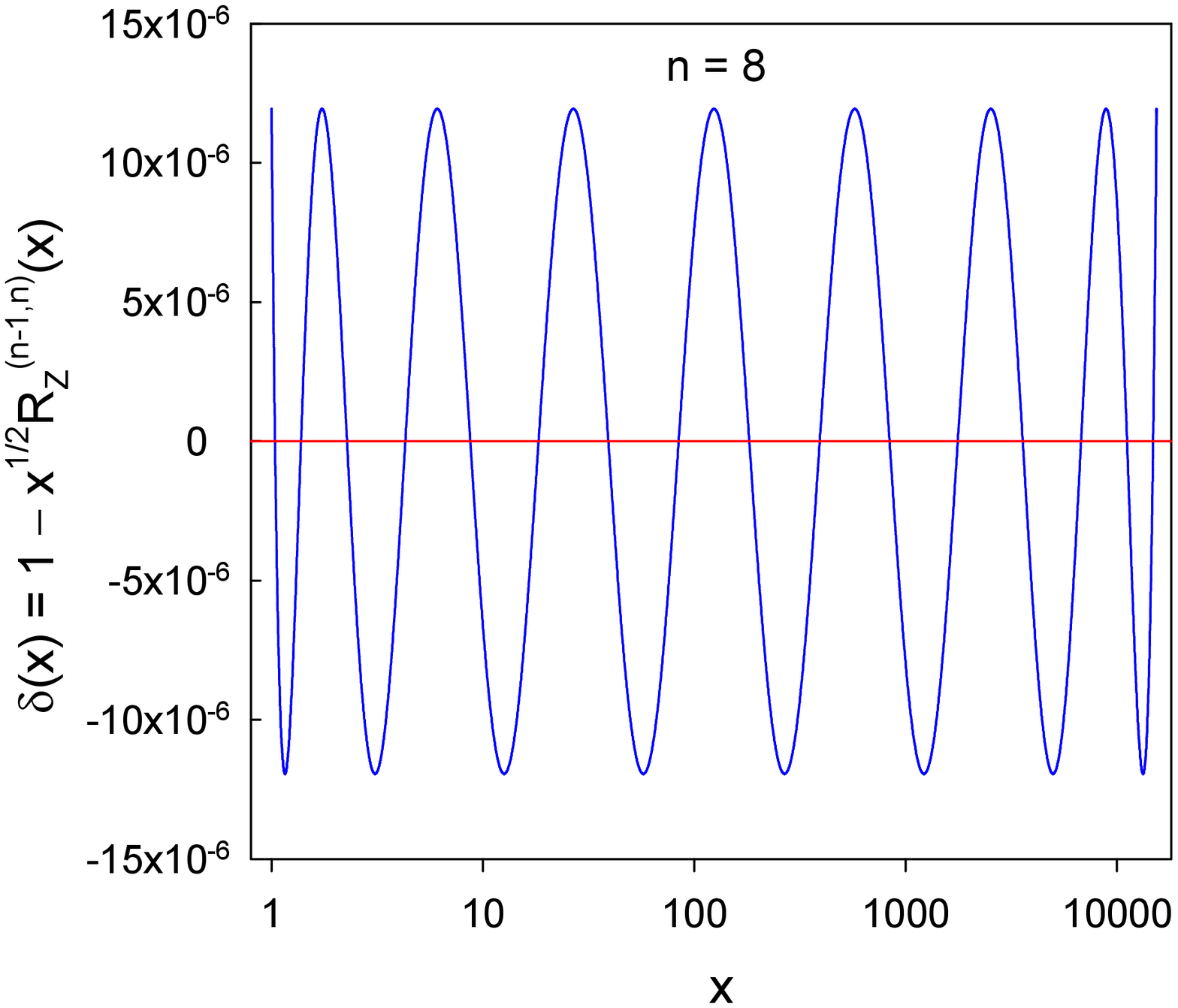}
&
\includegraphics*[width=8cm,clip=true]{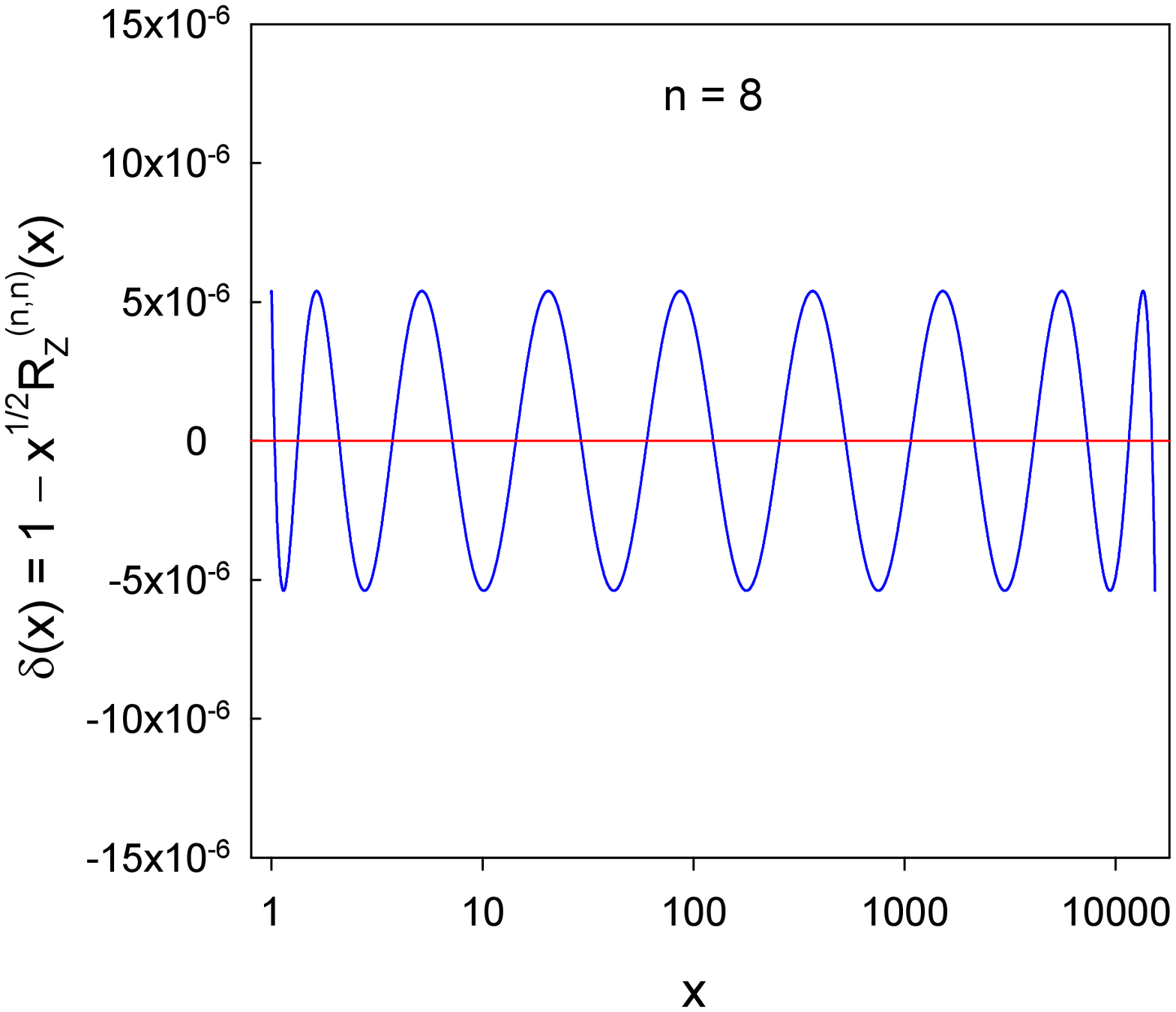}
\\ (a) & (b)
\end{tabular}
\caption{The deviation $ \delta(x) = 1 - \sqrt{x} R_Z^{(m,n)}(x) $ of the Zolotarev optimal rational polynomial $ R_Z^{(m,n)}(x) $ 
         for $ x \in [1,b] $, $ b = \lambda_{max}^2/\lambda_{min}^2 = (6.20/0.05)^2 = 15376 $. 
         In the left figure (a), $ (m,n) = (n-1,n)=(7,8) $, while in the right figure (b), $ (m,n)=(n,n)=(8,8) $.
         In both cases, $ \delta(x) $ has $ (m+n+2) $ alternate change of sign in the 
         interval $ [1,b] $, with $ (m+1) $ minima and $ (n+1) $ maxima, all of the same magnitude, satisfying 
         the necessary and sufficient condition of the optimal rational approximation of $ 1/\sqrt{x} $ 
         for $ x \in [1, b] $, acccording to de la Vall\'{e}e-Poussin's theorem and Chebycheff's theorem.}
\label{fig:zolo_lmin005_lmax620}
\end{center}
\end{figure*}

In Fig. \ref{fig:zolo_lmin005_lmax620}, we plot the deviation $ \delta^{(m,n)}_Z (x) = 1 - \sqrt{x} R_Z^{(m,n)}(x) $ 
of the Zolotarev optimal rational polynomials $ R_Z^{(7,8)}(x) $ and $ R_Z^{(8,8)} $ for $ x \in [1,b] $,    
$ b = \lambda_{max}^2/\lambda_{min}^2 = (6.20/0.05)^2 = 15376 $. 
In both cases, $ \delta(x) $ has $ (m+n+2) $ alternate change of sign in the 
interval $ [1,b] $, with $ (m+1) $ minima and $ (n+1) $ maxima, all of the same magnitude, thus satisfying 
the necessary and sufficient condition for the optimal rational approximation of $ 1/\sqrt{x} $ 
in the interval $ [1, b] $, acccording to de la Vall\'{e}e-Poussin's theorem and Chebycheff's theorem.
Note that $ d_Z \simeq 1.19447 \times 10^{-5} $ in the left figure (a),  
which is about 2.2 times of $ d_z \simeq 5.39351 \times 10^{-6} $ in the right figure (b), i.e.,  
$ d_Z^{(n-1,n)} \simeq 2.2 d_z^{(n,n)} $, as pointed out in Ref. \cite{Chiu:2002eh}.

\subsection{$N_s = 2 n $ (even)}

For the case $ N_s = 2 n $ (even), the requirement of $ R_5 $ symmetry implies that $ \omega_s = \omega_{N_s+1- s} $, and 
\bea
\label{eq:delta_2n}
\delta(\lambda) = 1 - S(\lambda) 
                = \frac{ 2 \prod_{s=1}^{n} (1-\omega_s \lambda)^2}
                       {\prod_{s=1}^{n}(1-\omega_s \lambda)^2 + \prod_{s=1}^{n} (1+\omega_s \lambda)^2} \ge 0,  
\eea
unlike the $ \delta_Z^{(n-1,n)}(x) $ (shown in Fig. \ref{fig:zolo_lmin005_lmax620}(a)) which varies alternatively 
between $ +d_Z $ and $ -d_Z $. 
Nevertheless, one can shift $ \delta_Z(x) $ by a constant $ +d_Z $, simply by changing the overall coefficient $ D_0 $ 
of $ R_Z^{(n-1,n)} $ (\ref{eq:zolo_even}) to $ D_0'$,  
\bea
\label{eq:D_0'}
D_0' = \frac{1}{\sqrt{\xi}} \frac{ \prod_{l=1}^n ( \xi +C_{2l-1}) }{ \prod_{l=1}^{n-1} ( \xi +C_{2l} ) }, \hspace{4mm} 
       \xi = \frac{1}{1-{\kappa'}^2 \mbox{sn}^2 \left( \frac{K'}{2n}; \kappa' \right) }. 
\eea
Then the maximal deviation becomes $ 2 d_Z $, which is twice of that of the Zoloterev optimal rational polynomial.
Obviously, this is the optimal chiral symmetry one can have for DWF with $ R_5 $ symmetry.  
This can be attained by requiring $ \delta(\lambda) $ to have $ 2n+1 $ alternative maxima and minima for 
$ \lambda \in [\lambda_{min}, \lambda_{max}] $,  
\bea
\label{eq:dZS_(n-1,n)}
\delta(\lambda) = 2 d_Z, \ 0, \ \cdots, \ 2d_Z, \ 0, \ 2d_Z,  \hspace{6mm}  d_Z = \frac{1-\Lambda}{1+\Lambda},  
\eea
at the consecutive points ($ \lambda_{min} = \lambda_1 < \lambda_2 < \cdots < \lambda_{2n+1} = \lambda_{max} $), 
where $ d_Z $ is the maximum deviation $ | 1- \sqrt{x} R_Z^{(n-1,n)}(x) |_{\rm max} $ of the 
Zolotarev optimal rational polynomial $ R_Z^{(n-1,n)}(x) $ for $ 1/\sqrt{x} $, $ x \in [1, b] $.
This immediately implies that the weights in (\ref{eq:D_odwf}) can be fixed by the positions of the minima of $ \delta(x) $,      
\bea
\label{eq:omega_sym_even}
\omega_s = \omega_{N_s+1-s} = \frac{1}{\lambda_{min}} \sqrt{ 1-{\kappa'}^2 \mbox{sn}^2 \left( \frac{(2s-1) K'}{N_s}; \kappa' \right) }, 
\hspace{4mm} s = 1, \cdots, N_s/2,  
\eea 
where $ \mbox{sn}(u ; \kappa') $ is the Jacobian elliptic function with modulus $ \kappa' = \sqrt{1-1/b} $, 
and $ K' $ is the complete elliptic function of the first kind with modulus $ \kappa' $.
Then the approximate sign function $ S(H) $ of the effective 4-dimensional Dirac operator (\ref{eq:D_eff}) becomes 
\bea 
S(H) 
     &=& h D'_0 \frac{ \prod_{l=1}^{n-1} (h^2 + C_{2l})}{\prod_{l=1}^{n} (h^2 + C_{2l-1})}, \hspace{4mm} h = \frac{H}{\lambda_{min}}, 
\eea
which is exactly the same as the Zolotarev optimal rational approximation of the sign function (see Eq. (\ref{eq:zolo_even})) 
except replacing $ D_0 $ by $ D_0' $.   
Note that if we replace $ D_0 $ by $ D_0^{''} $, 
\bea
\label{eq:D_0''}
D_0^{''} = \frac{ \prod_{l=1}^n ( 1 +C_{2l-1}) }{ \prod_{l=1}^{n-1} ( 1+C_{2l} ) }, 
\eea
then $ \delta(x) $ is shifted by a constant $ -d_Z $, thus its maxima become $ 0 $, and minima $ -2 d_Z $.
However, this cannot be realized by DWF with even $ N_s $ and $ R_5 $ symmetry, due to the constraint (\ref{eq:delta_2n}),  
$ \delta(x) \ge 0 $.   

\begin{figure*}[tb]
\begin{center}
\includegraphics*[width=8cm,clip=true]{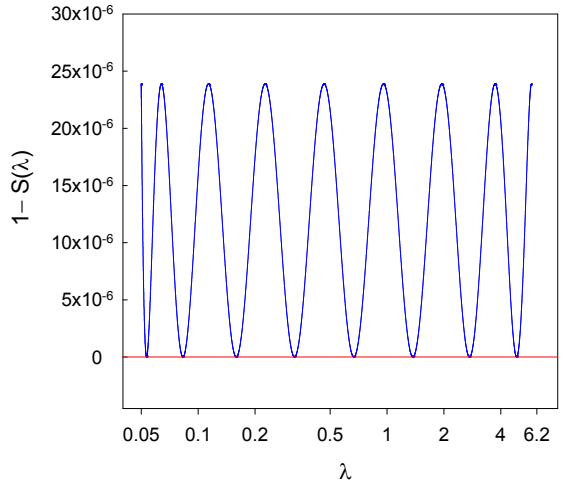}
\caption{The deviation $ 1 - S(\lambda) $ of the optimal DWF with $ R_5 $ symmetry, 
         for $ N_s = 16 $, $ \lambda_{max}/\lambda_{min} = 6.20/0.05 $,  
         where the weights are computed according to (\ref{eq:omega_sym_even}). 
}
\label{fig:odwf_Ns16_delta_positive}
\end{center}
\end{figure*}

In Fig. \ref{fig:odwf_Ns16_delta_positive}, we plot $ 1- S(x) $ of the optimal DWF with $ R_5 $ symmetry, 
for $ N_s = 16 $, $ \lambda_{max}/\lambda_{min} = 6.20/0.05 $, and the weights according to (\ref{eq:omega_sym_even}). 
We see that $ 1-S(x) $ has $ 2n+1=17 $ alternate maxima and minima in the interval $ [\lambda_{min}, \lambda_{max}] $, 
with $ 8 $ minima and $ 9 $ maxima, and $ |1-S(\lambda)|_{max} = 2 d_Z $, where $ d_Z $ 
is the maximum deviation $ | 1-\sqrt{x} R_Z^{(7,8)}|_{max} $ of the Zolotarev optimal rational polynomial, 
as shown in Fig. \ref{fig:zolo_lmin005_lmax620}(a).  

\subsection{$N_s = 2 n +1 $ (odd)}

For the case $ N_s = 2n + 1 $, the requirement of $ R_5 $ symmetry implies that $ \omega_s = \omega_{N_s+1- s} $, and 
\BAN
\delta(\lambda) = 1 - S(\lambda) = \frac{ 2 (1-\omega_{n+1} \lambda ) \prod_{s=1}^{n} (1-\omega_s \lambda)^2}
     {(1-\omega_{n+1} \lambda) \prod_{s=1}^{n}(1-\omega_s \lambda)^2 + (1+\omega_{n+1} \lambda) \prod_{s=1}^{n} (1+\omega_s \lambda)^2},  
\EAN
which can be positive, zero, or negative. 
Thus, we have two options to attain the optimal chiral symmetry with $ R_5 $ symmetry, i.e., to shift $ \delta(\lambda) $
by a constant $ +d_z $, or $-d_z$. 

The first option is to require $ \delta(\lambda) $ to have $ 2n+2 $ alternative maxima and minima 
in the interval $ [\lambda_{min}, \lambda_{max}] $,  
\BAN
\delta(\lambda) =  2 d_z, \ 0, \cdots, \, 2 d_z, \ 0, \hspace{4mm}  d_z = \frac{1-\sigma}{1+\sigma},  
\EAN
at consecutive points ($ \lambda_{min} = \lambda_1 < \lambda_2 < \cdots < \lambda_{2n+2} = \lambda_{max} $),  
where $ d_z $ is the maximum deviation $ | 1- \sqrt{x} R_Z^{(n,n)}(x) |_{\rm max} $ of the 
Zolotarev optimal rational polynomial $ R_Z^{(n,n)}(x) $ for $ 1/\sqrt{x} $ in the interval $ [1, b] $.
Thus the weights in (\ref{eq:D_odwf}) can be fixed by the positions of the minima of $ \delta(x) $,     
\bea
\label{eq:omega_sym_odd_positive}
\omega_s = \omega_{N_s+1-s} = \frac{1}{\lambda_{min}} \sqrt{ 1-{\kappa'}^2 \mbox{sn}^2 \left( \frac{(2s-1) K'}{N_s}; \kappa' \right) }, 
\hspace{4mm} s = 1, \cdots, (N_s+1)/2.  
\eea 
Note that the smallest weight $ \omega_{n+1} $ is chosen to be the unpaired one at the center of the fifth dimension, 
such that $ \delta(\lambda) $ does not change sign for all $ \lambda \in [ \lambda_{min}, \lambda_{max}] $.
Then the approximate sign function $ S(H) $ of the effective 4-dimensional Dirac operator (\ref{eq:D_eff}) becomes 
\bea 
S(H) 
     &=& h d_0' \frac{ \prod_{l=1}^{n} (h^2 + c_{2l})}{\prod_{l=1}^{n} (h^2 + c_{2l-1})}, \hspace{4mm} h = \frac{H}{\lambda_{min}}, 
\eea
which is exactly the same as the Zolotarev optimal rational approximation of the sign function (see Eq. (\ref{eq:zolo_odd})) 
except for the overall coefficent $ d_0' $,  
\bea
\label{eq:d_0'}
d_0' = \frac{1}{\sqrt{\xi}} \prod_{l=1}^n \frac{ ( \xi +c_{2l-1}) }{( \xi +c_{2l} ) }, \hspace{4mm} 
                            \xi = \frac{1}{1-{\kappa'}^2 \mbox{sn}^2 \left( \frac{K'}{2n+1}; \kappa' \right) }. 
\eea

The second option is to require $ \delta(\lambda) $ to have $ 2n+2 $ alternative maxima and minima 
in the interval $ [\lambda_{min}, \lambda_{max}] $,  
\BAN
\delta(\lambda) =  0, \ -2 d_z, \cdots, \, 0, \ -2 d_z, \hspace{4mm}  d_z = \frac{1-\sigma}{1+\sigma},  
\EAN
at consecutive points ($ \lambda_{min} = \lambda_1 < \lambda_2 < \cdots < \lambda_{2n+2} = \lambda_{max} $).  
This immediately implies that the weights in (\ref{eq:D_odwf}) can be fixed by the positions of the maxima of $ \delta(x) $,     
\bea
\label{eq:omega_sym_odd_negative}
\omega_s = \omega_{N_s+1-s} = \frac{1}{\lambda_{min}} \sqrt{1-{\kappa'}^2 \mbox{sn}^2 \left( \frac{(N_s+3-2s) K'}{2N_s}; \kappa' \right) }, 
\hspace{4mm} s = 1, \cdots, (N_s+1)/2.
\eea 
Note that the largest weight $ \omega_{n+1} $ is chosen to be the unpaired one at the center of the fifth dimension, 
such that $ \delta(\lambda) $ does not change sign for all $ \lambda \in [ \lambda_{min}, \lambda_{max}] $.
Then the approximate sign function $ S(H) $ of the effective 4-dimensional Dirac operator (\ref{eq:D_eff}) becomes 
\bea 
S(H) 
     &=& h d_0^{''} \prod_{l=1}^{n} \frac{(h^2 + c_{2l})}{(h^2 + c_{2l-1})}, \hspace{4mm} h = \frac{H}{\lambda_{min}}, 
\eea
which is exactly the same as the Zolotarev optimal rational approximation of the sign function (see Eq. (\ref{eq:zolo_odd})) 
except for the overall coefficient $ d_0^{''} $,  
\bea
\label{eq:d_0''}
d_0^{''} = \prod_{l=1}^n \frac{ ( 1 +c_{2l-1}) }{( 1 +c_{2l} ) }. 
\eea
          
\begin{figure*}[tb]
\begin{center}
\begin{tabular}{@{}c@{}c@{}}
\includegraphics*[width=8cm,clip=true]{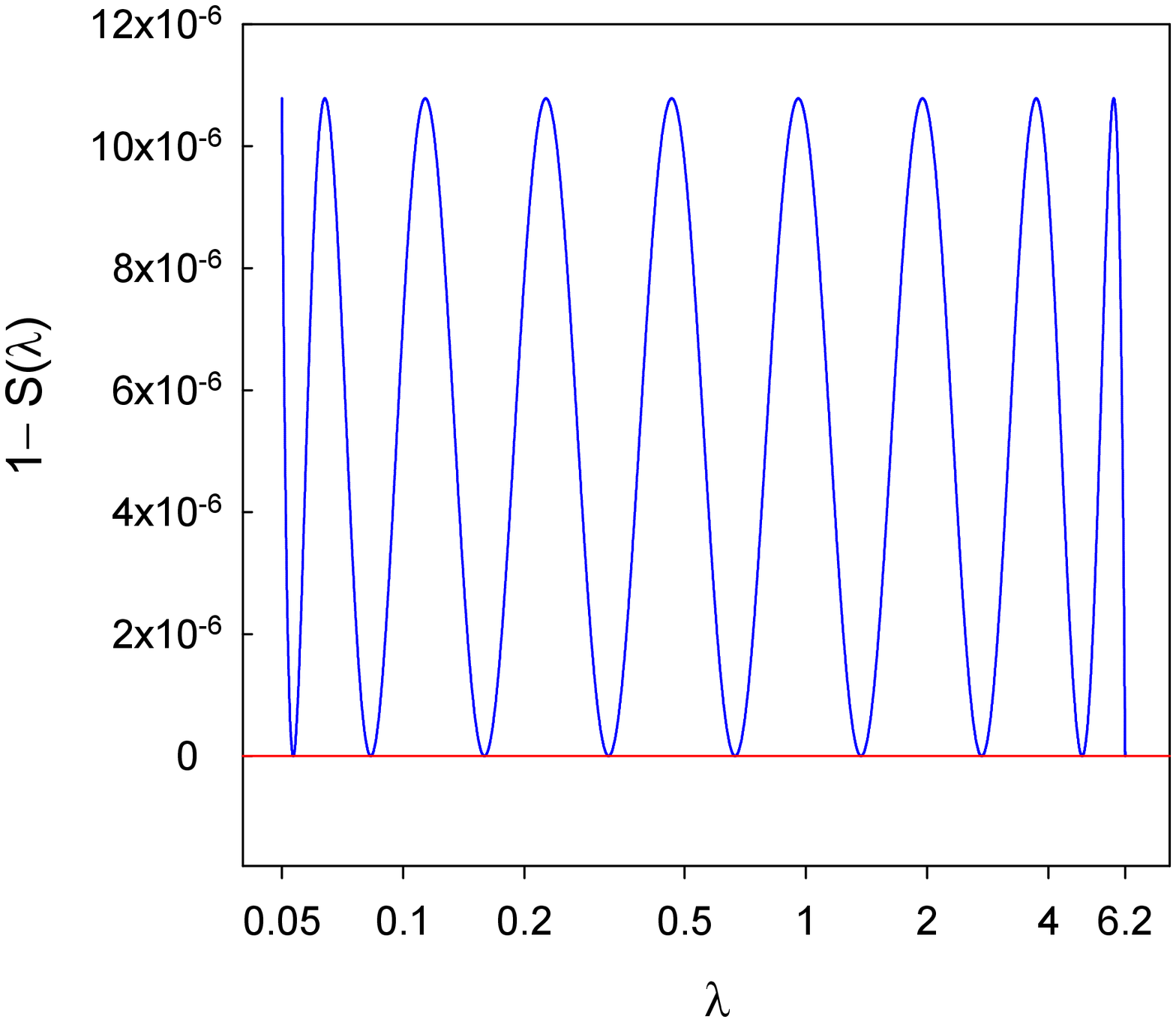}
&
\includegraphics*[width=8cm,clip=true]{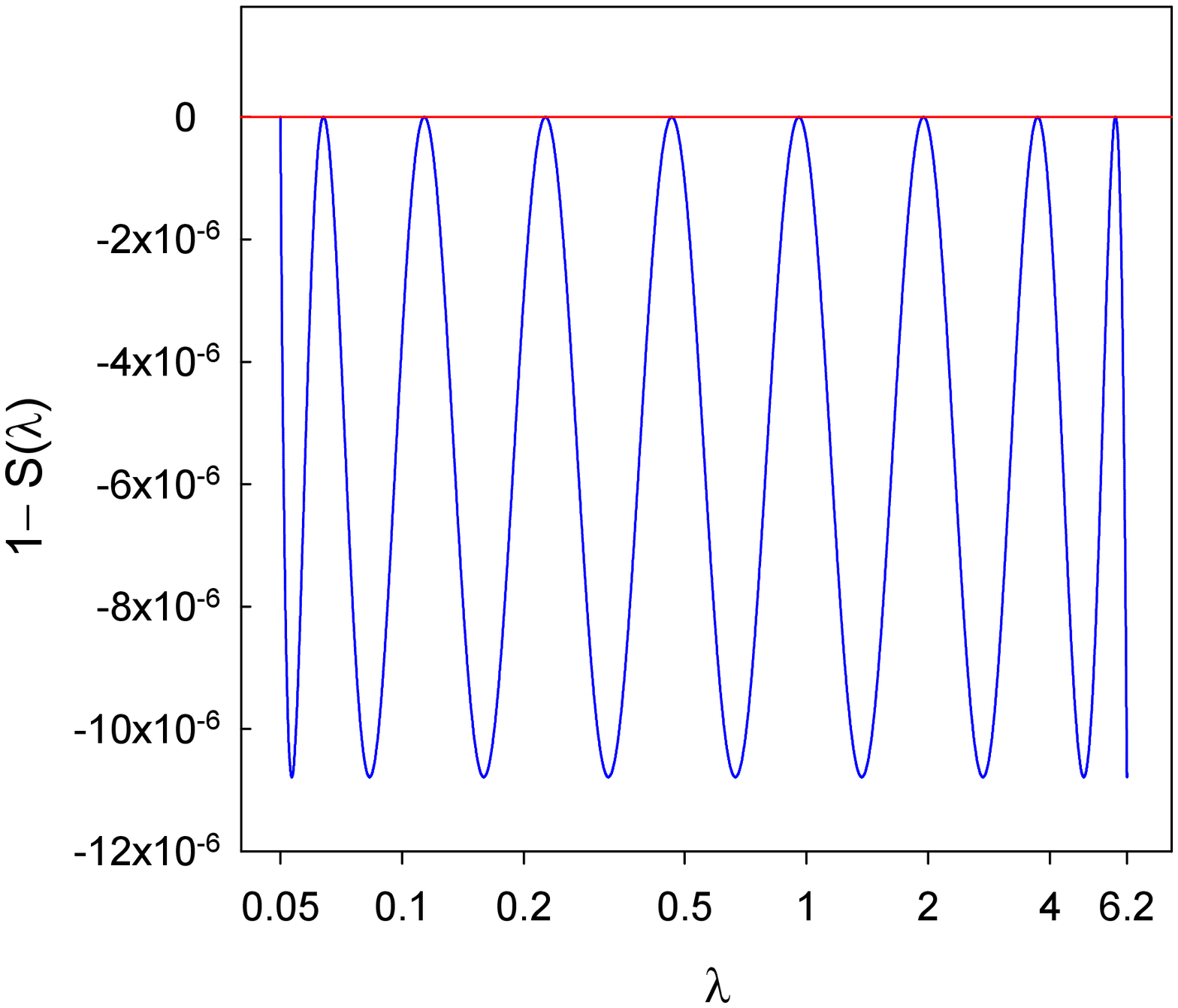}
\\ (a) & (b)
\end{tabular}
\caption{The deviation $ 1 - S(\lambda) $ of the optimal DWF with $ R_5 $ symmetry, 
         for $ N_s = 17 $, $ \lambda_{min} = 0.05 $, and $ \lambda_{max} = 6.20 $.  
         In (a), the weights are computed according to (\ref{eq:omega_sym_odd_positive}), 
         while in (b), according to (\ref{eq:omega_sym_odd_negative}).} 
\label{fig:odwf_Ns17}
\end{center}
\end{figure*}

In Fig. \ref{fig:odwf_Ns17}, we plot $ 1- S(x) $ of the optimal DWF with $ R_5 $ symmetry, 
for $ N_s = 17 $, and $ \lambda_{max}/\lambda_{min} = 6.20/0.05 $. 
In Fig. \ref{fig:odwf_Ns17}(a), the weights are computed according to (\ref{eq:omega_sym_odd_positive}), 
while in Fig. \ref{fig:odwf_Ns17}(b), according to (\ref{eq:omega_sym_odd_negative}). 
In both cases, $ 1-S(\lambda) $ has $ 2n+2 = 18 $ alternate maxima and minima in the interval $ [\lambda_{min}, \lambda_{max}] $, 
with $ 9 $ minima and $ 9 $ maxima. The maximum deviation $ |1-S(\lambda) |_{max} = 2 d_z $, 
where $ d_z $ is the maximum deviation $ | 1-\sqrt{x} R_Z^{(8,8)}|_{max} $ of the Zolotarev optimal rational polynomial, 
as shown in Fig. \ref{fig:zolo_lmin005_lmax620}(b).

\section{Numerical Test}

Theoretically, the sign function error $ |1-S(H)| $ of the optimal DWF with $ R_5 $ symmetry is twice of that without $ R_5 $ symmetry.
Nevertheless, it is interesting to check their difference in large-scale simulations of lattice QCD,    
by computing the residual mass \cite{Chen:2012jya}
\BAN
\label{eq:mres}
m_{res} &=& \frac{1}{4r} \frac{ \left< \tr\{ [D^{-1}(m_q)]^{\dagger}(1-S^2) D^{-1}(m_q) \}_{0,0} \right>_U}
                              { \left< \tr\{ [(D_c + m_q)^{-1}]^{\dagger} (D_c + m_q)^{-1} \}_{0,0} \right>_U}, \\
        &=& \frac{\left< \tr(D_c + m_q)^{-1}_{0,0} \right>_U}
                 {\left< \tr[\gamma_5 (D_c + m_q) \gamma_5 (D_c+m_q)]^{-1}_{0,0} \right>_U} - m_q,
\EAN
where $ (D_c + m_q)^{-1} $ denotes the valence quark propagator,
tr denotes the trace running over the color and Dirac indices, and the brackets $ \left< \cdots \right>_U $
denote averaging over an ensemble of gauge configurations. 


To generate the gauge ensemble, 
we perform the hybrid Monte Carlo simulation of $(2+1)$-flavors QCD on the $24^3 \times 48$ lattice with the Wilson gauge action
at $ \beta = 6/g^2 = 6.10$ (lattice spacing $ a \sim 0.06 $~fm), for the sea-quark masses $ m_u a=m_d a= 0.005 $ and $ m_s a = 0.04 $,  
with pion mass $ \sim 260 $ MeV. 
For the quark part, we use optimal DWF with $ R_5 $ symmetry for $ u $, $d $ and $ s $ quarks, 
with $ c = 1, d = 0 $ (i.e., $ H = H_w $), $ N_s = 16 $, and $ \lambda_{max}/\lambda_{min} = 6.20/0.05 $. 
The weights are computed according to the formula (\ref{eq:omega_sym_even}). 
The strange quark is simulated with the exact pseudofermion action for one-flavor DWF \cite{Chen:2014hyy}, 
and the up and down quarks are simulated with the 2-flavors algorithm as outlined in \cite{Chen:2014hva}.
We generate the initial 400 trajectories with 2 Nvidia GTX-TITAN GPUs working together. 
After discarding the initial 240 trajectories for thermalization, we sample one configuration every 5 trajectories, 
resulting 32 ``seed" configurations.
Then we use these 32 seed configurations as the initial configurations for 32 independent simulations on 32 Nvidia GTX-TITAN GPUs.
Each GPU generates 100-110 trajectories, and we accumulate a total of 3300 trajectories.
From the saturation of the binning error of the plaquette, as well as the evolution of the topological charge,
we estimate the autocorrelation time to be around 10 trajectories. Thus we sample one configuration every 10 trajectories,
and obtain $330$ configurations for this ensemble.

Then we compute two sets of quark propagators with point source at the origin and $ m_q a = 0.005 $, 
one set with the $ \{ \omega_s \} $ exactly the same as the sea-quarks (with $ R_5 $ symmetry), 
and the other set with the $ \{ \omega_s \} $ (\ref{eq:omega_s}) without $ R_5 $ symmetry. 
The residual mass of the set with $ R_5 $ symmetry is $ (m_{res} a)^{R_5} = 0.00024(1)$, 
while that of the set without $ R_5 $ symmetry is $m_{res}a = 0.00014(1)$. 
Thus the ratio $  (m_{res})^{R_5}/m_{res} \simeq 1.71(14) $, consistent with the theoretical expectation.


\section{Concluding Remarks}

With the weights (\ref{eq:omega_sym_even}), (\ref{eq:omega_sym_odd_positive}) and (\ref{eq:omega_sym_odd_negative}) for 
optimal DWF with $ R_5 $ symmetry, together with those (\ref{eq:omega_s}) for optimal DWF without $ R_5 $ symmetry \cite{Chiu:2002ir},  
this completes the study of DWF with the approximate sign function $ S(H) $ in the effective 4-dimensional lattice Dirac operator 
satisfying $ | 1 - S(\lambda) | \le 2 d_Z $ (with $ R_5 $ symmetry) or $ | 1 - S(\lambda) | \le d_Z $ (without $ R_5 $ symmetry) 
for $ \lambda^2 \in [\lambda_{min}^2, \lambda_{max}^2] $,  
where $ d_Z $ is the maximum deviation $ | 1- \sqrt{x} R_Z(x) |_{\rm max} $ of the 
Zolotarev optimal rational polynomial $ R_Z(x) $ of $ 1/\sqrt{x} $ for $ x \in [1, \lambda_{max}^2/\lambda_{min}^2] $,  
with degrees $ (n-1,n) $ for $ N_s = 2n $, and $ (n,n) $ for $ N_s = 2 n + 1 $.

The correspondence between the approximate sign function $ S(H) $ of the optimal DWF and the Zolotarev optimal 
rational polynomial gives the optimal rational approximation a more general viewpoint.
In general, the deviation $ \delta(x) = 1 - \sqrt{x} R_Z(x) $ can be shifted by a constant $ \epsilon $, 
simply by adjusting the overall coefficient $ D_0 $ in (\ref{eq:zolo_even}) or $ d_0 $ in (\ref{eq:zolo_odd}). 
In particular, for $ R_Z^{(n-1,n)} $, if $ D_0 $ is replaced by $ D_0' $ (\ref{eq:D_0'}), then $ \epsilon = +d_Z $, 
$ \delta_{max} = 2 d_Z $ and $ \delta_{min} = 0 $; while by $ D_0^{''} $ (\ref{eq:D_0''}), $ \epsilon = -d_Z $, 
$ \delta_{max} = 0 $ and $ \delta_{min} = -2 d_Z $.
Similarly, for $ R_Z^{(n,n)} $, if $ d_0 $ is replaced by $ d_0' $ (\ref{eq:d_0'}), $ \epsilon = +d_z $, 
$ \delta_{max} = 2 d_z $ and $ \delta_{min} = 0 $; while by $ d_0^{''} $ (\ref{eq:d_0'}), $ \epsilon = -d_z $, 
$ \delta_{max} = 0 $ and $ \delta_{min} = -2 d_z $.
Even though this does not satisfy the criterion that the maxima and the minima of $ \delta(x) $ all have the same magnitude 
and $ \delta_{min} = -\delta_{max} $, the most salient featues of the optimal rational approximation are preserved, namely, 
the number of alternate maxima and minima is $(m+n+2)$, with $ (n+1) $ maxima and $ (m+1) $ minima, 
as well as all maxima (minma) are equal. 
We can regard this as the generalized optimal rational approximation (with a constant shift $ \epsilon $).
For $ \epsilon > 0 $, this can be realized by DWF with/without $ R_5 $ symmetry. 
However, for $ \epsilon < 0 $, it cannot be realized by DWF with $ R_5 $ symmetry and $ N_s = 2n $ (even), 
due to the constraint (\ref{eq:delta_2n}), $ \delta(\lambda) \ge 0 $.

\section*{Appendix}

In this appendix, we collect the basic formulas of the Zolotarev optimal rational polynomials $ R_Z^{(n-1,n)}(x) $ and 
$ R_Z^{(n,n)}(x) $ for the inverse square root function $ 1/\sqrt{x} $, $ x \in [1, b] $. 
\bea
\label{eq:zolo_even}
R_Z^{(n-1,n)}(x) &=& D_0 \frac{ \prod_{l=1}^{n-1} (x + C_{2l})}{\prod_{l=1}^{n} (x + C_{2l-1})}
                  =  \sum_{l=1}^{n} \frac{B_l}{x + C_{2l-1} }, \\
\label{eq:zolo_odd}
R_Z^{(n,n)}(x) &=& d_0 \prod_{l=1}^{n} \frac{x + c_{2l}}{x + c_{2l-1}}  
                =  (x + c_{2n}) \sum_{l=1}^{n} \frac{b_l}{x + c_{2l-1}}, 
\eea
where
\BAN
&&  C_l = \frac{\mbox{sn}^2(\frac{lK'}{2n}; \kappa' ) } {1-\mbox{sn}^2(\frac{lK'}{2n}; \kappa' )},  \hspace{4mm}
    c_l = \frac{\mbox{sn}^2(\frac{lK'}{2n+1}; \kappa' ) } {1-\mbox{sn}^2(\frac{lK'}{2n+1}; \kappa' )}, \\
&&  B_l = D_0 \frac{ \prod_{i=1}^{n-1} ( C_{2i} - C_{2l-1} ) } { \prod_{i=1, i \ne l}^{n} ( C_{2i-1} - C_{2l-1} ) }, \hspace{4mm}
    b_l = d_0 \frac{ \prod_{i=1}^{n-1} ( c_{2i} - c_{2l-1} ) } { \prod_{i=1, i \ne l}^{n} ( c_{2i-1} - c_{2l-1} ) }, \\
&&  D_0 = \frac{2 \Lambda }{1+ \Lambda } \frac{ \prod_{l=1}^n (1+C_{2l-1}) }{ \prod_{l=1}^{n-1} ( 1+C_{2l} ) }, \hspace{4mm}
    d_0 = \frac{2 \sigma }{1+ \sigma} \prod_{l=1}^n \frac{1+c_{2l-1}}{1+c_{2l}}, \\
&&  \Lambda = \prod_{l=1}^{2n} \frac{\Theta^2 \left(\frac{2lK'}{2n};\kappa' \right)}
                                    {\Theta^2 \left(\frac{(2l-1)K'}{2n};\kappa' \right)}, \hspace{4mm}
    \sigma = \prod_{l=1}^{2n+1} \frac{\Theta^2 \left(\frac{2lK'}{2n+1};\kappa' \right)}
                                     {\Theta^2 \left(\frac{(2l-1)K'}{2n+1};\kappa' \right)}.
\EAN
Here $ \mbox{sn}(u ;\kappa') = \eta $ is the Jacobian elliptic function with modulus $ \kappa' = \sqrt{1-1/b} $, 
as defined by the integral
\BAN
\label{eq:sn}
u(\eta) = \int_{0}^\eta \frac{dt}{\sqrt{(1-t^2)(1-{\kappa'}^2 t^2 )}},  
\EAN
$ K'=u(1) $ is the complete elliptic function of the first kind with modulus $ \kappa' $,
and $ \Theta $ is the elliptic theta function.

For $ R_Z^{(n-1,n)} $, the deviation $ \delta(x) = 1 - \sqrt{x} R_Z^{(n-1,n)}(x) $  
has $ 2n+1 $ alternate change of sign in the interval $ [1, b] $,
and attains its maxima and minima,
\BAN
\label{eq:dZ_(n-1,n)}
\delta(x) =  d_Z, \ -d_Z, \ \cdots, \ d_Z, \ -d_Z, \ d_Z,  \hspace{6mm}  d_Z = \frac{1-\Lambda}{1+\Lambda},  
\EAN
at consecutive points ($ 1 = x_1 < x_2 < \cdots < x_{2n+1} = b $), 
\BAN
\label{eq:x_(n-1,n)}
x_i = \frac{1}{1-{\kappa'}^2 \mbox{sn}^2 \left( \frac{(i-1) K'}{2n}; \kappa' \right) }, \hspace{2mm} i = 1, \cdots, 2n+1. 
\EAN
 
For $ R_Z^{(n,n)} $, the deviation $ \delta(x) = 1 - \sqrt{x} R_Z^{(n,n)}(x) $  
has $ 2n+2 $ alternate change of sign in the interval $ [1, b] $,
and attains its maxima and minima,
\BAN
\label{eq:dZ_(n,n)}
\delta(x) =  d_z, \ -d_z, \ \cdots, \ d_z, \ -d_z,  \hspace{6mm}   d_z = \frac{1-\sigma}{1+\sigma}, 
\EAN
at consecutive points ($ 1 = x_1 < x_2 < \cdots < x_{2n+2} = b $), 
\BAN
\label{eq:x_(n,n)}
x_i = \frac{1}{1-{\kappa'}^2 \mbox{sn}^2 \left( \frac{(i-1) K'}{2n+1}; \kappa' \right) }, \hspace{2mm} i = 1, \cdots, 2n+2.
\EAN

\begin{acknowledgments}

  This work is supported in part by the Taiwan Ministry of Science and Technology 
  (No.~NSC102-2112-M-002-019-MY3) and NTU-CQSE (Nos.~NTU-ERP-103R891404, NTU-ERP-104R891404).
  The manuscript was completed during a visit at the University of Kentucky. 
  The author would like to thank Keh-Fei Liu for his kind hospitality and support.

\end{acknowledgments}


\end{document}